\documentclass[twocolumn,preprintnumbers,amssymb,pre,superscriptaddress]{revtex4}

\usepackage{graphicx}
\usepackage{dcolumn}
\usepackage{bm}
\usepackage{amstext}

\newcommand{\trp}[1]{{#1}} 
\newcommand{\hcf}[1]{{#1}}

\begin{document}

\bibliographystyle{unsrt}


\title{Beating patterns of filaments in viscoelastic fluids}
\author{Henry C. Fu}
\email{Henry_Fu@brown.edu}
\affiliation{Division of Engineering, Brown University, Providence, RI 02912}
\author{Charles W. Wolgemuth}%
\email{cwolgemuth@uchc.edu}
\affiliation{Department of Cell Biology, University of Connecticut Health Center, Farmington, CT 06030}%
\author{Thomas R. Powers}
\email{Thomas_Powers@brown.edu}
\affiliation{Division of Engineering, Brown University, Providence, RI 02912}

\date{\today}


\begin{abstract}
Many swimming microorganisms, such as bacteria and sperm, 
\trp{use flexible flagella to move through viscoelastic media
in their natural environments.}
In this paper we address
the effects a viscoelastic fluid has on the motion and beating
patterns of elastic filaments.  We treat both a passive filament which
is actuated at one end, and an active filament with bending forces
\trp{\hcf{arising from} internal motors distributed}
along its length.  We describe how viscoelasticity modifies the
hydrodynamic forces exerted on the filaments, and how these modified
forces affect the beating patterns.  We 
\trp{show}  how \trp{high viscosity} of purely viscous or viscoelastic solutions can 
lead to the experimentally observed beating patterns of
sperm flagella, 
in which motion is concentrated at the distal end of the flagella. 
\end{abstract}


\maketitle

\section{Introduction}

\begin{figure}
\includegraphics[height=0.7in]{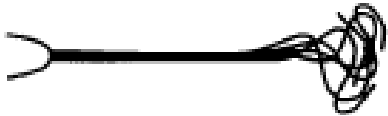}
\caption{\trp{The waveform of a human sperm observed by Ishijima et al.~\cite{Ishijima1986} in a solution with very high viscosity, 400\,cP. } } \label{OneWaveform}
\end{figure}

\trp{
Eukaryotes use beating cilia and flagella to transport fluid and swim~\cite{bray2001}. Cilia and flagella share a common structure, consisting of a core axoneme of nine doublet microtubules arranged around two inner microtubules. Molecular motors slide the microtubule doublets back and forth to generate the observed beating patterns. Cilia typically have an asymmetric beating pattern, with a power stoke in which the extended cilium pivots about its base, and a recovery stroke in which the cilium bends sharply as it returns to its position at the beginning of the cycle~\cite{Sleigh1974}. This stroke pattern is effective for moving fluid past the body of the cell, such as the epithelial cells of the airway or the surface of a swimming \textit{Paramecium}. On the other hand, flagella typically have a symmetric planar or helical waveform~\cite{Sleigh1974},  such as the flagellum of bull sperm which exhibits a traveling wave with an amplitude that increases with distance from the head of the sperm~\cite{Rikmenspoel1984}. The shape of the centerline of a beating filament determines the rate of transport or swimming. Many studies show that this shape depends on the properties of the medium. For example, the flagella of human sperm have a slightly helical waveform in water. As viscosity increases via the addition of polymers, the waveform becomes less helical and the amplitude flattens along most of the flagellum, with all the deflection taking place at the free end~\cite{Ishijima1986} (Fig.~\ref{OneWaveform}). Similar effects are observed in cervical mucus and other viscoelastic solutions~\cite{Rikmenspoel1984,SuarezDai1992}.}

\trp{In this paper, we calculate the shape of a beating filament as a function of the properties of the medium. Since the size scale for the filaments is tens of microns, and since typical swimming speeds are tens to hundreds of microns per second, the Reynolds number is very small, and inertia is unimportant. The medium is modeled with a single-relaxation time fading-memory model for a polymer solution~\cite{BirdArmstrongHassager1977}. We consider two different models for the filament. First, we consider the passive case, in which one end of a flexible filament is waved up and down~\cite{machin1958,wiggins_goldstein1998,YuLaugaHosoi2006,LaugaFloppy2007}.  We also consider a more realistic active model in which undulations arise from internal sliding forces distributed all along the length of the filament~\cite{camalet_et_al1999,Riedel-Kruse2007}.} Although 
active filaments are a closer representation of eukaryotic flagella, we
show that many of the important features of swimming filaments in
viscoelastic media are also present in the simple passive filament. \trp{Furthermore, passive filaments} may be more amenable to experiment and quantitative comparison with theory.
\trp{Our main results are} qualitative
explanations for some of the beating shapes observed in sperm in
high-viscosity and viscoelastic solutions~\cite{Ishijima1986,Rikmenspoel1984,Suarez_et_al1991,SuarezDai1992,HoSuarez2003}.  \trp{These results indicate that the observed shape change with increasing viscosity is a physical rather than a behavioral response. We reported some of our results for the case of an active filament in a viscoelastic medium with zero solvent viscosity in a previous publication~\cite{FuPowersWolgemuth2007}. In addition to presenting new results for the passive filament in a viscoelastic medium, in the present paper we also systematically study the dependence of the shape of an active filament on solvent viscosity. }

\trp{To make the analysis as simple as possible, we work to linear order in the deflection of the filament away from a straight configuration. Although the power dissipated by a beating filament is second order in deflection, it is sufficient to calculate the shape to first order in deflection to accurately compute the power to second order. We calculate how this power varies as the beating pattern changes with increasing relaxation time. There are also important changes in the swimming velocity due to shape change.}
Even in a linearly viscoelastic fluid, for which the
swimming speed of a filament with prescribed beating pattern is the
same as in a Newtonian fluid~\cite{FulfordKatzPowell1998}, changes to the beating patterns due to
the viscoelastic response can lead to changes in swimming velocities.
However, as \trp{has been previously emphasized}
~\cite{Lauga2007,FuPowersWolgemuth2007}, in viscoelastic fluids the swimming
velocity also receives corrections from the nonlinearity \trp{of the fluid constitutive relation} at second order,
and 
both these nonlinear corrections and the changes to the
beating patterns must be taken into account 
\trp{to correctly calculate the swimming velocity~\cite{FuPowersWolgemuth2007}.}

\trp{We begin our analysis by reviewing commonly used theories for the internal forces acting on passive and active filaments. Then we introduce the Oldroyd-B fluid, the fading-memory model we will use throughout this paper. After providing context with a brief discussion of resistive force theory for Newtonian viscous fluids, we discuss resistive force theory for small amplitude motions of a slender filament in an Oldroyd-B fluid. Using the balance of internal and external forces, we calculate the shape of a beating filament for both the passive and active cases.  We close with a discussion of the limitations of our analysis and the implications of our results. }

\section{Models}

\subsection{\trp{Internal forces acting on} flexible filaments}

\trp{The internal forces acting on a filament may be specified either by the energy as a function of the curvature or a constitutive relation for the moment acting on a cross-section of the filament. The two approaches are equivalent, and in this paper we use  the moment on a cross-section. }

\subsubsection{Passive Filament}

We first consider the passive elastic filament. \trp{Let $\mathbf{r}(s)$ denote the path of the centerline of an inextensible filament of length $L$. Then the moment $\mathbf{M}$ due to internal stresses acting on the cross-section of a bent filament is
\begin{equation}
\mathbf{M}=A\mathbf{r}'\times\mathbf{r}'',
\label{passsiveConRelat}
\end{equation}
where $A$ is the bending modulus and primes denote differentiation with respect to $s$~\cite{landau_lifshitz_elas}. Since we only consider planar filaments in this paper, it is convenient to introduce the right-handed orthonormal frame  $\{\hat\mathbf{e}_1,\hat\mathbf{e}_2,\hat\mathbf{z}\}$, where $\hat\mathbf{e}_1=\mathbf{r}'$ is the tangent vector of $\mathbf{r}(s)$, and $\hat\mathbf{z}$ is the outward
normal to the plane of the filament. Thus, $\kappa=\hat\mathbf{e}_2\cdot \hat\mathbf{e}'_1$ is the curvature of the the curve $\mathbf{r}(s)$. Note
that the sign of $\kappa$ is meaningful since $\hat\mathbf{e}_2$ does not flip at inflection points. }
Throughout this paper, we use a coordinate system for which
the straight filament lies in the $\hat{\bf  e}_x$ direction, and consider
only planar motion of the filament in the $\hat {\bf e}_x$-$\hat{\bf
  e}_y$
plane.  Further, we restrict ourselves to small displacements of the
filament, so that  ${\bf r}(s) \approx s \hat {\bf e}_x + h(s)
\hat{\bf e}_y$.  For such small displacements, we may approximate
$\kappa\approx  h''$ and $\mathbf{M}\approx\hat\mathbf{z}Ah''$.  
The elastic force per unit length  on the filament is found by moment balance, 
\begin{equation}
\mathbf{M}'+\hat\mathbf{e}_1\times\mathbf{F}=0,\label{Mbal}
\end{equation}
where $\mathbf{F}$ is the force acting on a cross-section at $s$. For
small displacements, the force per unit length due to internal
stresses is therefore $\mathbf{f}_\mathrm{int}=\mathbf{F}'\approx- A
h'''' \hat {\mathbf{e}}_y$.  \hcf{Eq.~(\ref{Mbal}) also allows a tangential
force that may be identified as a tension.  For small displacements,
the tension force is higher order in displacements than the normal
forces, and so we ignore it in this work~\cite{camalet-julicher00}.}
In the small Reynolds number flows associated with swimming cells, the
total force on each element of the filament must be zero,
$\mathbf{f}_{\mathrm{ext}} + \mathbf{f}_{\mathrm{int}} =0$.
As we shall see \trp{below, to determine the shape of the filament,}
 it is sufficient to consider 
\trp{external} forces which are
linear in the velocity of filament elements.  \trp{This conclusion is true} even in nonlinearly viscoelastic
fluids. \trp{Therefore} force balance
yields a fourth order linear differential equation to be solved for
the time-dependent filament shape.


The differential equation must be supplemented with appropriate boundary conditions.  For
the free end, we have
\begin{eqnarray}
-A h'''(L)  &=& 0\\
A h''(L)&=& 0,
\end{eqnarray}
corresponding to zero force and torque, respectively.  For the driven
end, we may consider prescribed angle,
\begin{eqnarray}
h'(0) &=& \epsilon \cos( \omega t)\\
h(0) &=& 0,
\end{eqnarray}
or prescribed force
\begin{eqnarray}
-A h'''(0) &=& \epsilon \cos( \omega t)\\
h(0) &=& 0.
\end{eqnarray}


\subsubsection{Active Filament}

\begin{figure}
\includegraphics[height=1in]{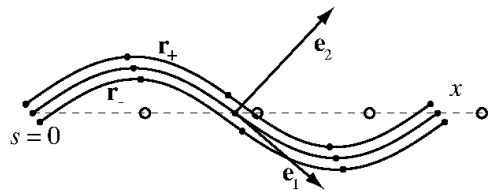}
\caption{The dotted line is the centerline of the flagellum when it is straight; the circles divide it in quarters. The solid lines represent the bent centerline and $\mathbf{r}_\pm$; the dots divide each line in quarters as measured along the respective contours. \trp{All three curves have length $L$.} } \label{ufig}
\end{figure}

For a simple model of a sperm flagellum which incorporates active
bending forces, we consider a swimmer consisting of a flagellum which
always lies in the plane, and 
\trp{disregard}
the presence of a head.  We model
the flagellum as two inextensible filaments of length $L$ 
which are parallel and have a
\trp{fixed, uniform} separation $a$.  The filaments can slide past
each other, which causes the flagellum to bend (Fig.~\ref{ufig}).  Define the paths of the two
filaments as
\begin{equation}
\mathbf{r}_\pm=\mathbf{r}\pm(a/2)\hat\mathbf{e}_2, \label{r1r2}
\end{equation}
\trp{where the frame $\{\hat\mathbf{e}_1,\hat\mathbf{e}_2,\hat\mathbf{z}\}$ again applies to the centerline $\mathbf{r}(s)$ of the filament (Fig.~\ref{ufig}).}
Equation~(\ref{r1r2}) implies $\mathrm{d}s_+- \mathrm{d}s_-= -a\kappa \mathrm{d}s$, \trp{where $s_\pm(s)$ is the arclength of $\mathbf{r}_\pm(s)$. For example, $\kappa<0$ in the interval $0<s<L/2$ of the curve $\mathbf{r}$ in Fig.~\ref{ufig}. Thus, the arclength $s_+(L/2)$  of $\mathbf{r}_+(s)$ is \textit{longer} than $L/2$; likewise, the arclength $s_-(L/2)$ is shorter than $L/2$. In the region $0<s<L/2$, the material points of $\mathbf{r}_+$ have slid backwards relative to the material points of $\mathbf{r}_-$. }
Integrating $\mathrm{d}s_+-\mathrm{d}s_-$, 
we find that the distance filament $\mathbf{r}_-$ slides past filament $\mathbf{r}_+$ at $s$ is
\begin{equation}
\Delta(s) = -\int_0^s a\kappa(s)\mathrm{d}s=-a\left[\theta(s)-\theta(0)\right],\label{Ds}
\end{equation}
where $\theta$ is the angle between $\mathbf{e}_1$ and $\hat{\bf e}_x$.  Note that Eq.~(\ref{Ds}) is exact, although from here on we assume
$a\kappa\ll1$.

The composite filament is actuated by molecular motors that slide $\mathbf{r}_\pm$ back and forth past each other. Let $f_\mathrm{m}$ denote the force per unit length in the $\hat\mathbf{e}_1$ direction that the upper filament $\mathbf{r}_+$ exerts on the lower filament $\mathbf{r}_-$ through the motors. To first order in deflection, the total moment acting across the cross-section at $s$ is therefore given by 
\begin{equation}
M_z=Ah''(s)+a\int_s^Lf_\mathrm{m}(s)\mathrm{d}s.
\label{Mz}
\end{equation}
In this model, the bending modulus $A$ is an effective modulus
representing the stiffness of the entire flagellar structure.  
We note that there are additional forces that may be included in
modeling eukaryotic flagella, such as those arising from linking
proteins that resist the sliding of filaments past each other.  In
this work we leave those out for simplicity; we have found that for
reasonable magnitudes of these forces they do not change our
results qualitatively. \trp{Moment balance, Eq.~(\ref{Mbal}), applies as before, leading to a force on the cross section at $s$ of 
$\mathbf{F}=(-Ah'''+af_\mathrm{m})\hat\mathbf{e}_y$,
or an internal force per unit length of $\mathbf{f}_\mathrm{int}=(-Ah''''+af_\mathrm{m}')\hat\mathbf{e}_y$.  }

The boundary conditions for the active filament involve extra terms
from the sliding force.  At the
free end, zero force  and moment  give 
\begin{eqnarray}
-Ah'''(L)+af_\mathrm{m}(L)&=&0\\
Ah''(L)&=&0.\label{FMbc}
\end{eqnarray}
At the attachment to the head we may consider a variety of
situations.  For a fixed head, we may consider clamped and hinged
boundary conditions.  For clamped boundary
conditions,
\begin{eqnarray}
h(0) &=& 0\\
h'(0) &=& 0.
\end{eqnarray}
For hinged boundary conditions where the total moment 
\trp{vanishes,}
\begin{eqnarray}
A h''(0) +a \int_0^L  f_{\mathrm{m}}(s) ds &=& 0\\
h(0) &=& 0.
\end{eqnarray}

\subsection{Linearly and nonlinearly viscoelastic fluids}
\trp{In the limit of zero Reynolds number,} the equation of motion of an incompressible medium is given by force balance, 
$-\nabla p +\nabla\cdot{\bm \tau} =0 $, where $p$ is the pressure
enforcing incompressibility, $\nabla \cdot \mathbf{v} = 0$; $\mathbf{v}$ is the velocity;
and $\bm \tau$ is the deviatoric stress tensor.

The material properties of the medium are given by the constitutive
relation, which specifies the stress in the medium as a function of
the strain and strain history of the material.  In a Newtonian fluid
with viscous forces, the stress tensor is 
proportional to the
rate of strain.   In contrast, in a viscoelastic fluid, there is an
additional elastic component to the response of the medium.  
This effect is incorporated by fading-memory models, in which the stress relaxes over time to the viscous stress.
We will focus on a particular fading-memory model, the Oldroyd-B fluid, which has the
constitutive relation
\begin{equation}
{\bm \tau} + \lambda \stackrel{\triangledown}{\bm\tau}= \eta \dot {\bm \gamma}
+ \lambda \eta_s  \stackrel{\triangledown}{\dot{\bm\gamma}} \label{OldroydB}.
\end{equation}
In this equation,  $\lambda$ is the single relaxation time, $ \stackrel{\triangledown}{\bm\tau} =
\partial_t {\bm \tau} + \mathbf v \! \cdot \! \nabla \bm \tau -
{(\nabla \mathbf v)}^{\mathsf{T}} \! \! \cdot \! \bm \tau - \bm \tau \! \cdot \!
\nabla \mathbf v$ is the upper-convected time derivative of ${\bm \tau}$, $\mathbf{v}$ is the velocity, $\eta$ is the polymer viscosity, $\dot{\bm\gamma}=\nabla{\mathbf v}+{(\nabla \mathbf v)}^{\mathsf{T}}$
is the strain rate, \trp{and $ \stackrel{\triangledown}{\dot{\bm\gamma}}$ is the upper-convected time derivative of the strain rate.}   
In this paper we will use a dot to denote
time differentiation.  The total viscosity $\eta = \eta_p +
\eta_s$ is the sum of polymer and solvent viscosities.  
The Oldroyd-B fluid is one of the simpler nonlinear constitutive
relations for a viscoelastic fluid.  The upper-convected time
derivative
is the source of
nonlinearities~\cite{BirdArmstrongHassager1977}.  
If the solvent viscosity is ignored ($\eta_s = 0$
), the Oldroyd-B fluid reduces to the Upper Convected Maxwell fluid, and if
there is no polymer ($\eta = \eta_s$) the Oldroyd-B fluid reduces to a
Newtonian fluid.  The Oldroyd B fluid is  known to describe elastic
and first normal stress effects as long as elongational flows are
not large; however, it does not capture shear-thinning or yield-stress
behaviors observed in some non-Newtonian fluids.

Although we have introduced the nonlinearly viscoelastic fluid, in
what follows we will only work to linear order in the 
displacements of filaments.  In that case, our results will be the
same as if we consider the same model with ordinary time derivatives,
which corresponds to a linear Maxwell-Kelvin fluid.

\section{Filaments in Newtonian fluids}

The motion and swimming properties of filaments in Newtonian fluids have been
actively studied by many researchers~\cite{machin1958,wiggins_goldstein1998,
  camalet_et_al1999, LaugaFloppy2007}.   \trp{Here we briefly summarize
results in the Newtonian case} which will be useful for
comparison to filaments in viscoelastic fluids.  
 
A common approach to calculating the viscous forces acting on a
flagellum is resistive force theory, a local drag theory in which the
forces per unit length acting on a thin filament are proportional to
the fluid velocity relative to the
filament~\cite{gray_hancock1955}. Although resistive force theory does
not completely capture the effects of hydrodynamic interactions,
macroscopic-scale experiments have shown that it is surprisingly
accurate for single
filaments~\cite{koehler_powers2000,YuLaugaHosoi2006,QianPowersBreuer2008}. For a
purely viscous liquid, the force per unit length acting on a filament
moving \trp{in an otherwise quiescent fluid}
is given by 
\begin{equation}
\mathbf{f}_{\mathrm{vis}}= - \zeta_{||}(\dot\mathbf{r})_{||} - \zeta_\perp(\dot\mathbf{r})_\perp,\label{rft}
\end{equation}
the subscripts $||$ and $\perp$ denote the directions along and perpendicular to the filament at $\mathbf{r}(s,t)$, respectively.  The
friction coefficients are proportional to the viscosity $\eta$,
with $\zeta_\perp\approx2\zeta_{||}$.

For simplicity, assume that there is only one frequency of motion for
the filament, so the displacement of the filament can be written
$h(s,t) = \mathrm{Re} \lbrace \tilde h(s,t) \exp(-\mathrm{i} \omega t) \rbrace
$.  In this case, using the resistive force theory, the $y$-component of
the equation of
motion for the passive and active filament are, respectively,
\begin{eqnarray}
 -\mathrm{i} \omega \zeta_\perp \tilde h &=& - A \tilde h''''\\
-\mathrm{i} \omega\zeta_\perp \tilde h &=& -A \tilde h''''+ a f_\mathrm{int}'.
\end{eqnarray}
The solutions of these can be obtained in a manner similar to that
described in Sec. \ref{ResultsSection} to give $h(s,t)$.  

From $h(s,t)$, 
we may calculate the power dissipated by the filament and the
swimming velocity of the filament.
At each material element of the filament, the power dissipated is the
inner product of the hydrodynamic force and the velocity.  The total
time-averaged power dissipated is therefore
\begin{equation}
 \langle P \rangle = \frac{\zeta_{\perp} \omega^2}{2} \int_0^L ds \; |
 \tilde h(s) |^2   
\end{equation}

The swimming velocity can be determined by the constraint under Stokes
flow that the total force on the swimmer is zero.  To lowest order,
the component of force in the $\hat{\bf{e}}_y$ direction is zero as a
result of the equation of motion for the filament shape.  The beating
motion of the filament produces a $\hat{\bf{e}}_x$ component of the
time-averaged force \trp{that} 
is second order in the deflection, 
which is balanced by a drag force
from overall translation, i.e. swimming with velocity $U_{\mathrm{RFT}}$ in the $\hat{\bf{e}}_x$ direction:
\begin{eqnarray}
U_{\mathrm{RFT}}  &=& \frac{ \omega (\zeta_\perp -
   \zeta_\parallel)}{2 L \zeta_\parallel} \mathrm{Im}  \int_0^L ds \;
   \tilde h' \tilde h^*
\label{Newtonianswimspeed}
\end{eqnarray}
In this paper we ignore the drag from the head, which would contribute
an additional \trp{factor in the expression for $U_\mathrm{RFT}$.}
Note
that for clamped boundary conditions, the torque exerted on the
filament \trp{at $s=0$ has a reaction torque that
causes the head and ultimately the whole sperm to rotate}
~\cite{LaugaFloppy2007}.  Here we ignore these
effects for two reasons: 1) Experiments looking at the shape of beating
filaments can be performed with an immobilized head.  2)
For swimming, calculations ignoring the head are simpler and can be
easily compared to previous work, e.g.~\cite{camalet_et_al1999}, and
also are more easily extended and compared to the calculations we will
perform in the viscoelastic fluid.  

It is important to note that it is only
necessary to solve for the beating motion to first order in deflection
amplitude to obtain the correct leading order results for \trp{the power and the swimming speed,
even though these quantities are of second order in deflection.}

\section{Filaments in viscoelastic fluids}

Having seen how to analyze filaments in Newtonian fluids, we 
\trp{turn to} viscoelastic fluids.  In a
viscoelastic fluid, the hydrodynamic force felt by the filament is no
longer given by 
Eq.~(\ref{rft}).   Due to the change
in forces, the equations of motion for the filament are changed, so
that the beating patterns change.  Since both the beating patterns and
forces affect the power dissipated and swimming velocity, we expect
both of those to change as well.  \trp{To calculate the power dissipated by a filament beating in a viscoelastic medium to leading order in the deflection, we only need the velocity and the force to first order. Thus, it is sufficient to find the shape to first order. From Eq.~(\ref{Newtonianswimspeed}) we expect that a change in the first-order shape $h$ will change the swimming velocity; however, there are additional second order corrections that are present even for a waveform of prescribed shape~\cite{Lauga2007,FuPowersWolgemuth2007}.} Consistency requires including both of these
second order corrections~\cite{FuPowersWolgemuth2007}.  In this paper we concentrate on the effects
of viscoelasticity on beating patterns and power dissipated. 

\trp{The calculation of the force per unit length acting on a slender filament undergoing large deflections in a viscoelastic fluid is a daunting challenge. Since we consider small deflections, we may consider the simpler problem of the forces per unit length acting on a slightly perturbed infinitely long cylinder with a prescribed lateral or longitudinal traveling wave. Elsewhere we solve for the flow induced by these motions using the Oldroyd-B fluid and calculate the forces acting on the deformed 
cylinder~\cite{FuPowersWolgemuth2008Scallop}. We find that the force per unit length $\mathbf{f}_\mathrm{Oldroyd}$ obeys
\begin{equation}
\mathbf{f}_\mathrm{Olroyd}+\lambda_1
 \dot\mathbf{f}_\mathrm{Oldroyd}=\mathbf{f}_\mathrm{vis} +
 \frac{\eta_s}{\eta} \lambda \dot \mathbf{f}_\mathrm{vis} 
 \label{rftOldroyd}.
\end{equation}
}In the limit of \trp{zero}
solvent viscosity, $\eta_s = 0$, 
\trp{Eq.~(\ref{rftOldroyd})} is the same as the resistive force theory proposed by Fulford,
Katz, and Powell for the linear Maxwell
model~\cite{FulfordKatzPowell1998}.  \hcf{Eq. \ref{rftOldroyd} is valid for
slender filaments and to first order in small deflections.}

\subsection{Application to passive and active filaments}

To apply the above results to the passive and active filament
models, we use the hydrodynamic force $\mathbf{f}_\mathrm{Oldroyd}$ 
in the
equations of motion.  For the passive filament \trp{with a single frequency of motion, we find}
\begin{equation}
 -\mathrm{i} \omega \frac{1 -  \mathrm{i\, De}_2}{1 - \mathrm{i \,De}} \zeta_\perp h = - A h'''',\\
\end{equation}
where the Deborah numbers $\mathrm{De} = \omega \lambda$ and
$\mathrm{De}_2 = \mathrm{De} \eta_s/\eta$. 
\trp{For the active filament we must include the motor forcing:}
\begin{equation}
-\mathrm{i} \omega \frac{1 - \mathrm{i \,De}_2}{1 -  \mathrm{i\, De}} \zeta_\perp h = -Ah''''+ a f_\mathrm{m}'.
\end{equation}
The beating pattern is determined by solving these equations of
motion, and once the beating pattern is known, the power dissipated
and swimming velocity can be calculated.  

The power dissipated is calculated in the same way as for a Newtonian
fluid, except that now the force per unit length is specified by
Eq.~(\ref{rftOldroyd}), \trp{leading to}
\begin{equation}
\langle P \rangle = \frac{\zeta_{\perp} \omega^2}{2} \frac{1 +
 \mathrm{De} \mathrm{De}_2}{1 + \mathrm{De}^2} \int_0^L ds \; |
 \tilde h(s) |^2 .  \label{VEpower}
\end{equation}

As noted \trp{above,}
once the beating pattern 
affects the swimming velocity.  However, there is an additional
correction to the swimming velocity in the nonlinearly viscoelastic
Oldroyd-B fluid.  Instead of Eq. \ref{Newtonianswimspeed}, the correct
expression for the swimming speed is~\cite{FuPowersWolgemuth2007, FuPowersWolgemuth2008Scallop}
\begin{eqnarray}
U &=& \frac{1 + \mathrm{De} \mathrm{De}_2}{1 + \mathrm{De}^2}
\left\langle \int_0^L ds \; h'(s,t) \dot h(s,t) \right \rangle \\
&=& \frac{1 + \mathrm{De} \mathrm{De}_2}{1 + \mathrm{De}^2} U_{\mathrm{RFT}},
\end{eqnarray}
where the \trp{second} 
line follows from the ratio
$\zeta_\perp/\zeta_\parallel = 2$.  This result applies to an infinite
cylinder moving with beating pattern $h(s,t) = \mathrm{Re}\lbrace \tilde h(s)
  \exp(-i \omega t) \rbrace$.  To apply this result to the finite
  filament, we imagine periodically replicating the calculated beating
  pattern of the filament so that it becomes infinitely long.  
Note that due to the periodicity, the integration only needs to be
performed from $0$ to $L$.   This result ignores end
effects from the flow around the end of the finite filament.
We have mentioned the effects of beating patterns on swimming
velocity here \trp{for completeness}.  In the rest of the paper we focus on the changes to
beating patterns themselves and power dissipation rather than swimming velocity.

\section{Results}\label{ResultsSection}

\subsection{Non-dimensionalization}
We can non-dimensionalize the filament equations of motion and boundary
conditions by measuring lengths in terms of $L$, $f_{\mathrm m}$ in
terms of $A/a L^2$, and time in terms of $\omega^{-1}$.
Velocities are measured in terms of $L \omega$ and angular velocities
in terms of $\omega$.  For notational
simplicity, after scaling we use the same symbols for the new
quantities.  
The non-dimensional equation of motion for the passive filament \trp{ with a single frequency} is
\begin{equation}
-{\mathrm{i\, Sp^4}}\frac{1 -  \mathrm{i\, De}_2}{1 - \mathrm{i\, De}} \tilde h +\tilde h'''' = 0 \label{NDPassive}
\end{equation}
with boundary conditions
\begin{eqnarray}
- h'''(1) &=& 0 \\
h''(1) &=& 0\\
h(0) &=& 0 
\end{eqnarray}
\trp{and either of}
\begin{eqnarray}
h'(0) &=& \epsilon \cos(t) \;\;\;\;({\mathrm{prescribed\;angle}})\\
h'''(0)&=& \epsilon \cos(t) \;\;\;\;({\mathrm{prescribed\;force}}).
\end{eqnarray}
The dimensionless parameter ${\mathrm{Sp}}= L \left( \omega
\zeta_\perp / A \right)^{1/4}$, \trp{known as the Sperm number,}
involves the ratio of the
bending relaxation time of the filament to the period of the
traveling wave.  In what follows the forcing
$\epsilon$ is chosen small enough so that we remain safely in the
linear regime.

The non-dimensional equation of motion for the active filament with a fixed head \trp{and single frequency} becomes
\begin{equation}
-{\mathrm{i\, Sp^4}}\frac{1 -  \mathrm{i\, De}_2}{1 - \mathrm{i\, De}} \tilde h +\tilde h'''' - \tilde f'_{\mathrm
  m} = 0, \label{NDActive}\\
\end{equation}
\trp{where $f_\mathrm{m}=\mathrm{Re}\lbrace \tilde f_\mathrm{m}(s)
  \exp(-i t) \rbrace$. The boundary conditions for Eq.~(\ref{NDActive}) are}
\begin{eqnarray}
- h'''(1) + f_{\mathrm m}(1) &=& 0 \\
h''(1) &=& 0\\
h(0) &=& 0
\end{eqnarray}
\trp{and either of}
\begin{eqnarray}
h'(0) &=&0 \;\;\;\;({\mathrm {clamped}})\\
h''(0) + \int_0^1 f_{\mathrm{m}}(s) ds &=& 0 \;\;\;\;\;({\mathrm{hinged}}).
\end{eqnarray}

\subsection{Passive filament}

For the passive filament in sinusoidal steady state, we examine the
solutions to Eq.~(\ref{NDPassive}) which are
\begin{equation}
h_{\mathrm h}(s) =\sum_{j=1}^4C_j\exp(r_j s),\label{hhom}
\end{equation}
where the $r_j$ solve
\begin{equation}
-\mathrm{i\, Sp}^4 \frac{1 - \mathrm{i\, De}_2}{1-  \mathrm{i\, De}} + r_j^4 = 0.\label{reqn}
\end{equation}
The four coefficients $C_j$ are determined by the boundary
conditions, and the full time-dependent solution is then given by
\begin{equation}
h(s,t) = \mathrm{Re} \left\{ e^{-\mathrm{i} t} h_{\mathrm h}(s) \right\}
\end{equation}

From the form of $h_\mathrm{h}$ in Eq.~(\ref{hhom}), we can already make a
general comment about the expected form of beating patterns.  The
exponentials in Eq.~(\ref{hhom}) will lead to traveling or standing
waves.  Standing waves can only be produced if there are pairs of
$r_j$ that are complex conjugates.  This happens only in the elastic
limit ($\mathrm{De} \rightarrow \infty$) for which the $r_j$ are
proportional to $\pm (1 \pm i)$.  Therefore in the viscous case the
beating pattern will always have the form of a traveling wave, but as
$\mathrm{De}$ increases the beating pattern takes on more
characteristics of a standing wave.  

\begin{figure}
\includegraphics[width = 3.5in]{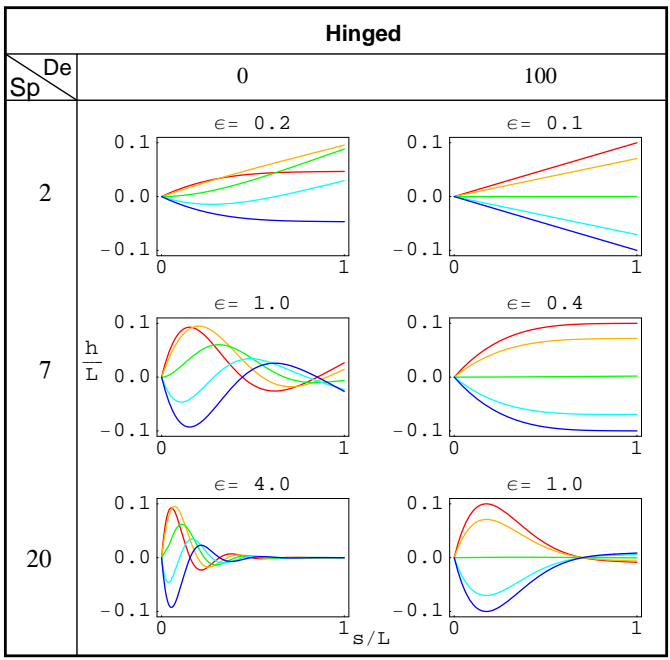}
\caption{(Color online.) Shapes of beating patterns for passive filaments driven by
  prescribed angle at the left end ($s=0$).  
  A half-cycle
  (red, orange, green, light blue, blue) of the pattern is shown for
  viscous ($\mathrm{De}=0$) and viscoelastic ($\mathrm{De}=100$) cases, and
  Sperm number $2$, $7$, and $20$. The ratio of solvent viscosity to the total viscosity is $\eta_\mathrm{s}/\eta=\mathrm{De_2}/\mathrm{De}=10^{-4}$.   At the
  top of each plot, the (dimensionless) angular magnitude $\epsilon$ required to produce motion with amplitude $0.1 L$ is shown.} \label{passiveshapes}
\end{figure}

Representative plots of these solutions for prescribed attachment
angles and forces are presented in Fig. \ref{passiveshapes}.
Solutions for the viscous ($\mathrm{De}=0$) and viscoelastic ($\mathrm{De}=100$) limits
are plotted side by side.  As expected, changing the viscoelastic
character of the fluid by varying the Deborah number induces
changes in the shapes of the beating patterns.  An important feature of these solutions is
the presence of a bending length scale $\xi$: $\xi/L =| r_j |^{-1} =
\mathrm{Sp}|(1 - \mathrm{i} \mathrm{De_2})/(1 - \mathrm{i} \mathrm{De}) |^{-1/4}$.   For $\mathrm{De}=0$, the purely
viscous case, this corresponds to the length scale set by the Sperm
number; a stiff filament with small Sperm number has large $\xi$ and
acts as a rigid rod, while a flexible filament with large \trp{Sperm} 
number has small $\xi$ and can bend in response to viscous forces.
Increasing the Deborah number acts to increase $\xi$.  
In the figure, the amplitude of the prescribed driving angle is
selected so that the maximum displacement of the filament is $L/10$,
and then printed above each figure.   The beating pattern is linear in
the amplitude of the driving angle.  Due to larger bending lengths,
$\xi$, in viscoelastic fluids, smaller driving angles are typically
required to produce the same amplitude motion.  

The beating patterns in Fig. \ref{passiveshapes} also demonstrate that 
viscoelastic effects alter the nature of the beating patterns as
expected from the form of Eq.~(\ref{hhom}):
for $\mathrm{De}=0$, we
obtain traveling wave beating patterns, while for large $\mathrm{De}$, in
the elastic limit, we obtain standing wave beating patterns.  

\begin{figure}
\includegraphics[width=3in]{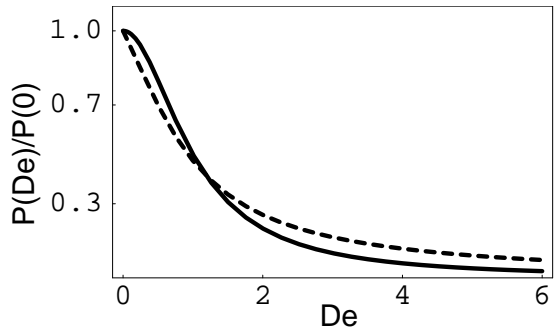}
\caption{Time-averaged power dissipated by passive filaments driven by
  prescribed angle, as a
  function of $\lambda = \omega \lambda$ (\trp{dashed curve}).  Power is
  normalized to the power dissipated in the viscous case $\mathrm{De}=0$.
  \trp{The black curve corresponds to the power dissipated if the beating pattern does not change from the viscous case, but the filament is placed in a viscoelastic medium.}  The driving amplitude
  is chosen so that the maximum amplitude of the beating pattern is
  $L/10$ in the viscous case. The ratio of solvent viscosity to the total viscosity is $\eta_\mathrm{s}/\eta=\mathrm{De_2}/\mathrm{De}=10^{-4}$.} \label{passivepower}
\end{figure}

From Eq.~(\ref{VEpower}) we obtain the 
time-averaged power dissipated 
\begin{eqnarray}
\langle P_M \rangle &=& \frac{\zeta_{\perp}}{2} \mathrm{Re} \left[ \int_0^L {\mathrm ds} \, 
    \omega^2 | \tilde h|^2\frac{ 1-  \mathrm{i\,De}_2}{1 -  \mathrm{i\,De}}  \right]\\
&=& \zeta_{\perp} \omega^2\frac{1 + \mathrm{De} \mathrm{De}_2}{2 (1 + \mathrm{De}^2)} \int_0^L
    {\mathrm d}s \, |\tilde h|^2\label{Pdiss}
\end{eqnarray}
Figure \ref{passivepower} 
\trp{shows} the power dissipation 
\trp{versus Deborah number (dashed curve)}. We also display the 
power dissipation, \trp{Eq.~(\ref{Pdiss}), evaluated with the purely viscous ($\mathrm{De}=0$) shape $\tilde h$ (solid curve). Note that the difference in shape $\tilde h$ between the viscous and viscoelastic cases has little effect; the power is mainly governed directly by the dependence of the force per unit length on $\mathrm{De}$.}

\subsection{Active Filaments}\label{activeshapes}

\subsubsection{Fixed Head}

\begin{figure*}
\includegraphics[width = 7in]{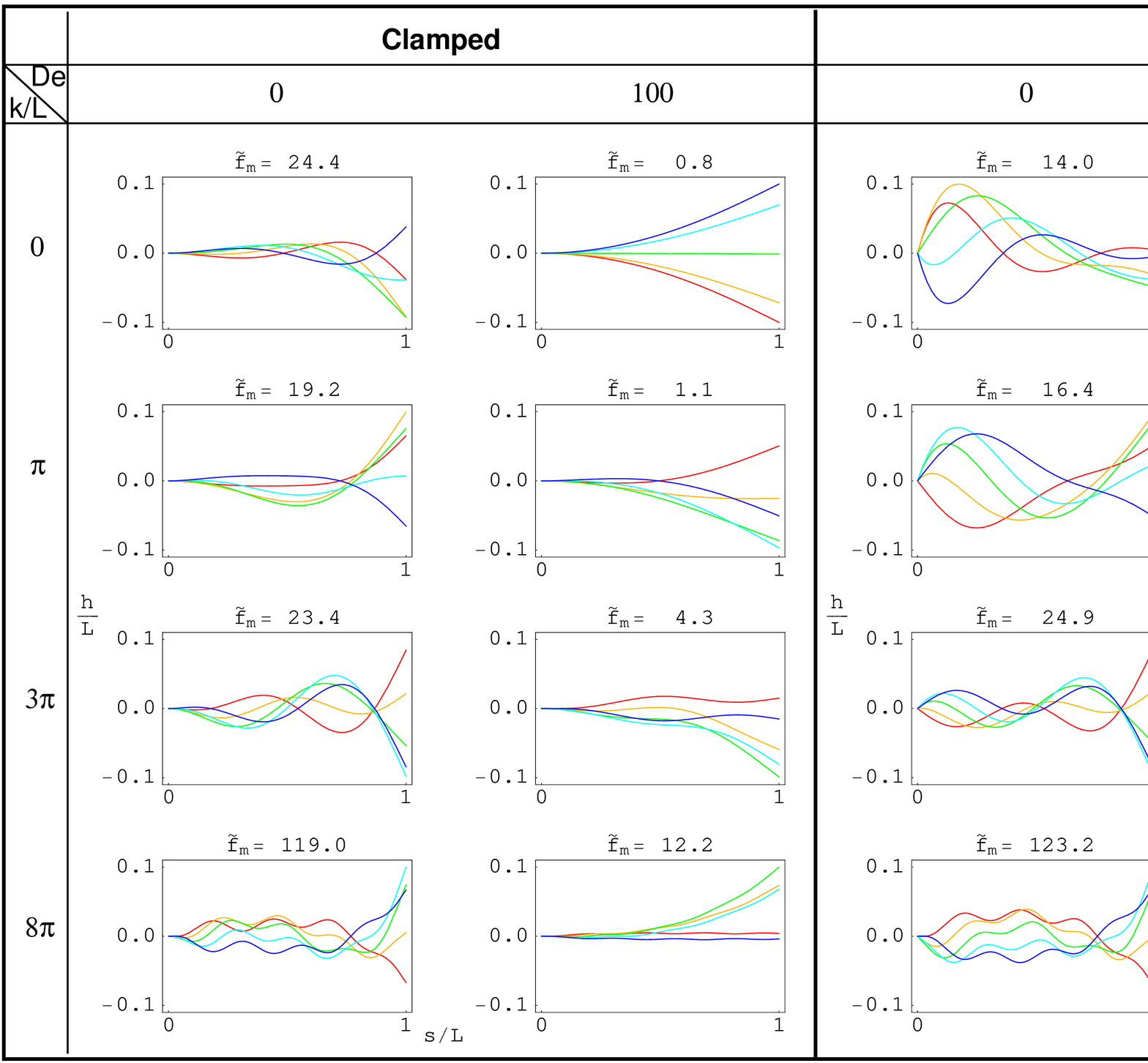}
\caption{(Color online.) Shapes of beating patterns for filaments with internal
  sliding forces and fixed head position, with $\mathrm{Sp} =7$.  A half-cycle
  (red, orange, green, light blue, blue) of the pattern is shown for
  viscous ($\mathrm{De}=0$) and viscoelastic ($\mathrm{De}=100$) cases, and
  for internal sliding forces given by Eq.~(\ref{waveforce}) with $k/L$
  varying from $0$ (uniform force) to $8 \pi$, as indicated.  At the
  top of each figure, the (dimensionless) magnitude $\tilde f_{\mathrm
  m}$ required to produce motion with amplitude $0.1 L$ is shown. The ratio of solvent viscosity to the total viscosity is $\eta_\mathrm{s}/\eta=\mathrm{De_2}/\mathrm{De}=10^{-4}$.} \label{fixedfig}
\end{figure*}

Examining Eq.~(\ref{NDActive}), we find that it only differs from the
passive case at the boundary conditions and by the presence of an
inhomogeneous term.  Therefore we may use the homogeneous solution
from the previous subsection,
\begin{equation}
h_{\mathrm{h}}(s) =\sum_{j=1}^4C_i\exp(r_j s)
\end{equation}
and add to it a particular solution.  As in the passive case,
beating patterns with traveling waves are expected in the viscous limit, and
increasing the Deborah number leads to beating patterns with
characteristics of standing waves.

To find the particular solution,
let us assume that the internal sliding force takes the form of a sliding wave
\begin{equation}
f_{\mathrm{m}}(s) = \mathrm{Re} \left\{ \tilde f_\mathrm{m} \exp\left(\mathrm{i} k
    s-\mathrm{i} t\right) \right\} \label{waveforce}
\end{equation} 
In that case, we can write the particular solution 
\begin{equation}
h_\mathrm{p}=\mathrm{i} k \tilde f_\mathrm{m} \mathrm{e}^{\mathrm{i}k s}
/\left[- \mathrm{i\, Sp}^4 (1-  \mathrm{i\,De}_2)/(1- \mathrm{i\,De}) + k^4 \right].
\end{equation}
Again the four boundary conditions set the four unknown
coefficients and the full time-dependent solution is
\begin{equation}
h(s,t) = \mathrm{Re} \left\{ e^{-\mathrm{i} t} \left[ h_{\mathrm h}(s) +
  h_{\mathrm{p}} \right] \right\}
\end{equation}

\begin{figure}
\includegraphics[width=3.3in]{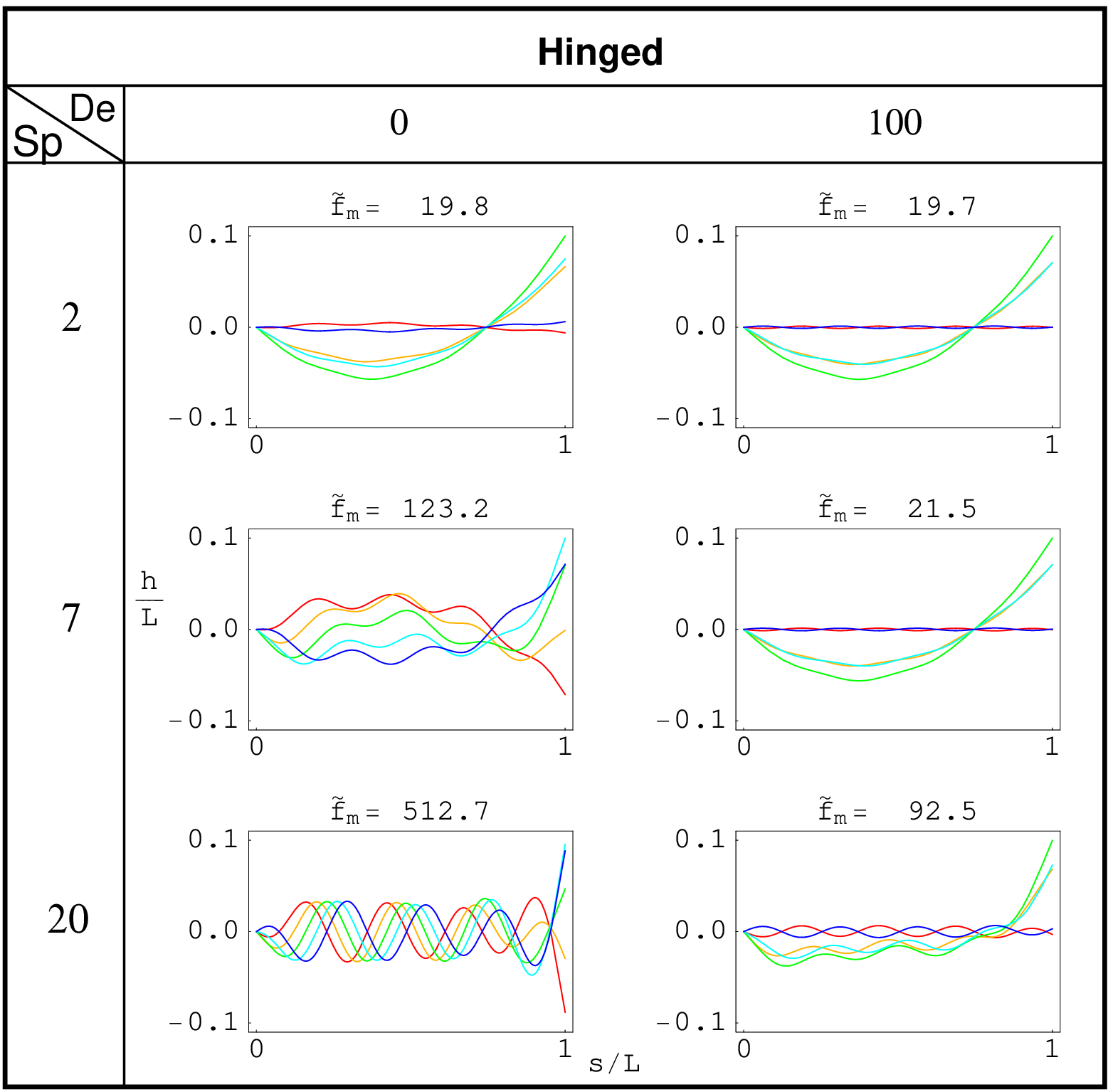}
\caption{(Color online.) Shapes of beating patterns for hinged filaments with internal
  sliding forces, with varying $\mathrm{Sp}$.  A half-cycle
  (red, orange, green, light blue, blue) of the pattern is shown for
  viscous ($\mathrm{De}=0$) and viscoelastic ($\mathrm{De}=100$) cases, and
  for internal sliding forces given by Eq.~\ref{waveforce} with $k/L=8
  \pi$.  At the top of each figure, the (dimensionless) magnitude
  $\tilde f_{\mathrm  m}$ required to produce motion with amplitude $0.1
  L$ is shown. The ratio of solvent viscosity to the total viscosity is $\eta_\mathrm{s}/\eta=\mathrm{De_2}/\mathrm{De}=10^{-4}$.} \label{Mafig}
\end{figure}

The beating shapes of the active flagellum and a fixed head are shown
in Fig.~\ref{fixedfig}, for $\mathrm{De} = 0$ (viscous case) and
$\mathrm{De}=100$ (viscoelastic case), and for various spatial dependencies of
the internal sliding force.  The forces required to produce beating
with amplitude $0.1 L$ are larger in the viscous cases than the
viscoelastic cases, \trp{again since the bending length $\xi$ is longer in the viscoelastic case than the viscous case.}
\trp{Just as in the case of the passive filament,} the effect of
the length scale $\xi$ can be  seen by comparing the viscous and
viscoelastic cases.  However, the wavelength  $2 \pi/k$ of the sliding force
introduces another length scale.  The combined presence of two
different length scales can be seen, for example, in the shape of the
hinged filament with $\mathrm{De}=0$, and \hcf{(dimensional)} $k=8 \pi/L$.  In this example, $\xi$ is
longer (and is the same as that seen for all the $\mathrm{De}=0$ shapes of
Fig.~\ref{fixedfig}), while the shorter scale $2 \pi/k$ leads to
ripples on top of the longer deformation.  Comparing the viscous to
viscoelastic case, one can also see 
\trp{a decrease in} the visible effects of the
shorter lengthscale arising from the internal forces 
as $\xi$ increases and the shape is dominated by drag forces.  
\trp{For example, consider} Fig.~\ref{Mafig},
which shows the effect of varying the Sperm number for an internal
sliding force with {(dimensional)} $k=8\pi/L$.  As the Sperm number increases, $\xi$
decreases and smaller lengthscale motions become more visible.  One implication of the presence of
these two length scales is that the observed ``wavelength'' of a
beating pattern does not necessarily give direct information about either
lengthscale.   

\begin{figure}
\includegraphics[width=2.5in]{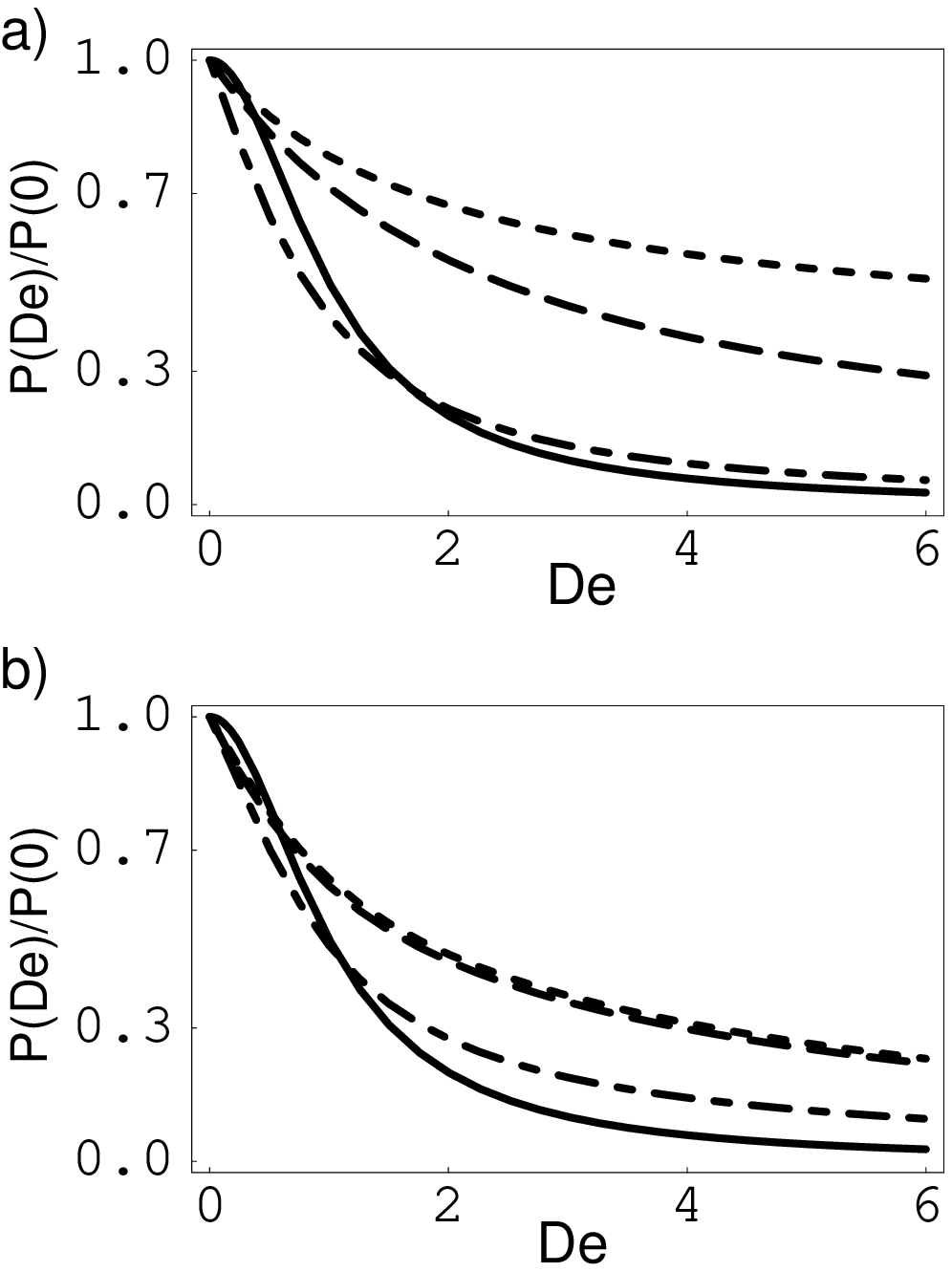}
\caption{Time-averaged power dissipated by filaments with active
  internal sliding forces, as a function of $\mathrm{De} = \omega \lambda$.  Power is
  normalized to the power dissipated in the viscous case $\mathrm{De}=0$,
  with force amplitude such that maximum filament displacement in the
  viscous case $\mathrm{De}=0$ is $0.1 L$.
  Different
  curves correspond to sliding forces given by Eq.~\ref{waveforce}
  with different values of $k$: dashes, $k=0$; long dashes, $k=\pi/L$;
  long dash-short dash, $k=3
  \pi/L$.  The solid curve is the power dissipated
  by a filament with a beating shape corresponding to the viscous case
  ($\mathrm{De}=0$).  The ratio of solvent viscosity to the total viscosity is $\eta_\mathrm{s}/\eta=\mathrm{De_2}/\mathrm{De}=10^{-4}$. a) Fixed head with clamped attachment point.  b)
  Fixed head with hinged attachment point.  } \label{powerfig}
\end{figure}

The time-averaged power dissipated by the clamped and hinged internally driven
filaments can be calculated as in the previous section.   
\trp{The power versus 
$\mathrm{De}$, normalized by
viscous power with $\lambda=0$, is shown in}
Fig.~\ref{powerfig}a and \ref{powerfig}b.  
Also shown for comparison is the power
dissipated  assuming that the beating shape of the viscous case is
prescribed for the viscoelastic cases.   For both prescribed forces and velocities, the growing
imaginary component of the viscosity shifts the drag force to be
out of phase with the velocity, decreasing the power dissipated.
However, for large $\mathrm{De}$ the power dissipated for prescribed forces is always
greater than that dissipated for prescribed velocities,
because as $\mathrm{De}$ increases, the real part of the drag decreases in
amplitude.  \trp{Thus,} prescribed forces lead to motion with larger
amplitude and velocity, and \trp{therefore} 
more power dissipated relative to the
case with prescribed velocities.  

\subsubsection{Comparison to experimental beating patterns}

\begin{figure*}
\includegraphics[width=16cm]{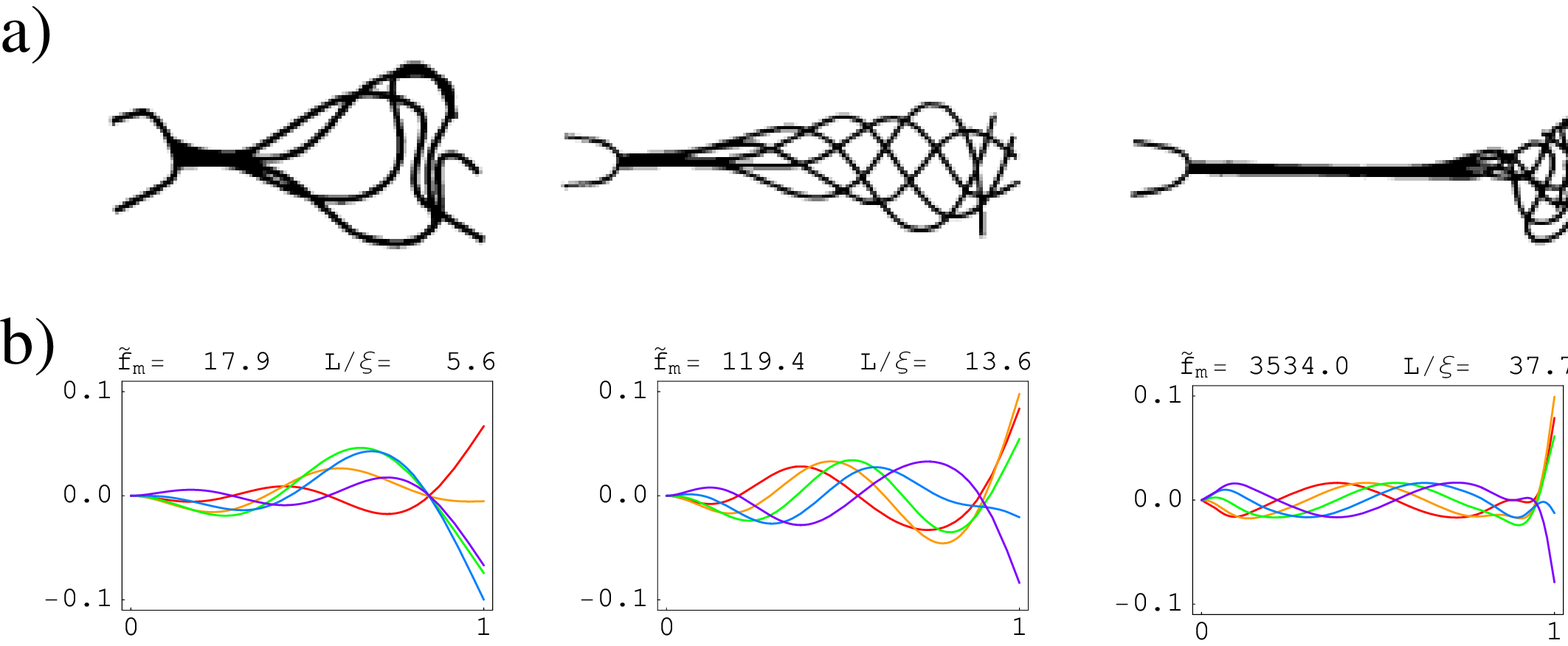}
\caption{(Color online.)  a)  Beating patterns of human sperm observed by Ishijima
   et al.~\cite{Ishijima1986}, in (from left to right) 1\,cP Hanks'
   solution, 35\,cP Hanks' solution, 4000\,cP Hanks' solution, 4360\,cP
   cervical mucus.   
b)  Shapes of beating patterns for filaments with internal
  sliding forces, fixed head position, and clamped boundary
  conditions.  A half-cycle
  (red, orange, green, light blue, blue) of the pattern is shown for
  internal sliding forces given by Eq.~\ref{waveforceA} with $k/L = 3 \pi$
  From left to right, the sperm numbers are 5.6, 13.6, 38.7, and 45,
  corresponding to viscosities of 1\,cP, 35\,cP, 4000\,cP, and 4360\,cP.
  In the first two plots, the medium is purely viscous, while in the
third plot, $\mathrm{De}=0.5$ and $\mathrm{De}_2 = 0.5/4000$. In the fourth plot, $\mathrm{De}=50$ and $\mathrm{De}_2 = 50/4360$.  
  At the top of each figure, the (dimensionless) magnitude $\tilde f_{\mathrm
  m}$ required to produce motion with amplitude $0.1 L$, and the
  lengthscale $\xi$ is shown.
   } \label{ExptShapes}
\end{figure*}

The equation of motion for the active filament (Eq.~\ref{NDActive}) allows us
to understand some of the qualitative changes in the shapes of beating
patterns of sperm flagella in different media.  There have been a
number of studies observing the different beating patterns of sperm in
media with increased viscosity and viscoelasticity~\cite{SuarezDai1992,
  Rikmenspoel1984, Ishijima1986}.  Of these, we focus on the paper of
Ishijima et al., because it combines a systematic study of beating patterns 
in
different media with measurements of viscosity using a falling-ball
rheometer, and because its use of a pipette to hold the heads of
observed sperm still makes its results amenable for modeling.  In Fig.~\ref{ExptShapes}a, the beating patterns of human sperm
in Hanks' medium with viscosity of 1\,cp, 35\,cp, 4000\,cp, and cervical
mucus with viscosity 4360\,cp are shown.   The 35\,cP and 4000\,cP
viscosity media are made by adding polyvinylpyrrolidone and
methylcellulose, \hcf{ respectively, to Hanks' medium (a Newtonian
  solution of salts and glucose in water)}.  
As viscosity increases, motion is concentrated at the distal end,
and motion in the  proximal and middle portions of the flagella is
damped out.  Similar behavior was
observed in Refs.~\cite{SuarezDai1992, Rikmenspoel1984}.  

At very high viscosity,
asymptotic analysis of the equation of motion Eq.~(\ref{NDActive}) demonstrates
that most of the beating motion 
\trp{occurs} at the end of the
flagellum.  For concreteness, because the beating
patterns in Fig.~\ref{ExptShapes}a were obtained while the sperm head was
held in place by a pipette, we use clamped boundary conditions to
model the motion.  Furthermore, since wavelengths of 26--28\,$\mu$m were
observed in the low viscosity media for flagella of overall length
$\approx 40 \mu$m, we assume a sliding force with \hcf{(nondimensional)} $k=3
\pi$ and the form of a traveling wave
\begin{equation}
f_{\mathrm{m}}(s) = \mathrm{Re} \left\{ \tilde f_\mathrm{m} e^{\mathrm{i} k
    s  - \mathrm{i}t} \right\} \label{waveforceA}
\end{equation} 
Then the inhomogeneous part of Eq.~(\ref{NDActive}) has the particular solution
\begin{equation}
h_\mathrm{p}=\mathrm{i} k \tilde f_\mathrm{m} \mathrm{e}^{\mathrm{i}k s}
/\left[- \mathrm{i\,Sp}^4 (1-  \mathrm{i\,De}_2)/(1- \mathrm{i\,De}) + k^4 \right].
\end{equation}
while the homogeneous part $h_\mathrm{h}(s)$ obeys
\begin{equation}
-{\mathrm{i\,Sp^4}}\frac{1 -  \mathrm{i\,De}_2}{1 -  \mathrm{i\,De}} h_\mathrm{h} + h_\mathrm{h}'''' = 0
\label{homogeneous}
\end{equation}
with boundary conditions
\begin{eqnarray}
h_\mathrm{h}'''(1) &=&  f_{\mathrm m}(1) - h_\mathrm{p}'''(1) \\
h_\mathrm{h}''(1) &=& -h_\mathrm{p}''(1)\\
h_\mathrm{h}(0) &=& -h_\mathrm{p}(0)\\
h_\mathrm{h}'(0) &=& -h_\mathrm{p}'(0).
\end{eqnarray}
We analyze the homogeneous solution by asymptotically matching
solutions appropriate for the middle of the flagellum with solutions appropriate for
``boundary layers'' at the ends of the flagellum. 
At high viscosity, the sperm number is large, and therefore the term
involving the fourth power of the sperm number in
Eq.~(\ref{homogeneous}) dominates.  Throwing out the other terms, we
find that the homogeneous solution vanishes.  This \trp{balance} is valid for the
middle of the flagellum, but in a boundary layer near the ends of the
flagellum, the homogeneous solution may
vary rapidly, making the term with derivatives important.  The
size of this boundary layer can be expected to be of order $\xi = L
|(1- \mathrm{i} \mathrm{De})(1- \mathrm{i\, De}_2)|^{1/4} /\mathrm{Sp} $.  
Measuring lengths by $\xi$, i.e. $x = s/ \xi$ for the proximal region, $x =
(L-s)/ \xi$ for the distal region, we obtain the equation of motion
\begin{eqnarray}
0&=&- \mathrm{e}^{\mathrm{i} \theta} h_\mathrm{h} + \partial_x^4 h_\mathrm{h}\\
\mathrm{e}^{\mathrm{i} \theta} &=& \frac{1- \mathrm{i\, De}_2}{1- \mathrm{i\,De}} \left|
\frac{1-  \mathrm{i\, De}}{1-  \mathrm{i\,De}_2} \right|
\end{eqnarray}
At the proximal end, the boundary conditions are
\begin{eqnarray}
h_\mathrm{h}(x=0) &=& - h_\mathrm{p}(0)\\
\partial_x h_\mathrm{h}(x=0) &=& - \partial_x h_\mathrm{p}(x=0)\\
h_\mathrm{h}(x \rightarrow \infty) &=& 0
\end{eqnarray}
While at the distal end the boundary conditions are
\begin{eqnarray}
\partial_x^3 h_\mathrm{h}(x=0) &=& \xi^3 f_\mathrm{m}(1) - \partial_x^3 h_\mathrm{p}(x=0)\\
\partial_x^2 h_\mathrm{h}(x=0) &=& - \partial_x^2 h_\mathrm{p}(x=0)\\
h_\mathrm{h}(x \rightarrow \infty) &=& 0\end{eqnarray}
In the boundary conditions, the last
constraint comes from the requirement that the solution in the boundary layers
asymptotically matches the solution in the middle of the flagellum.
The solution to these equations is a linear combination of four
exponential functions.
\begin{eqnarray}
h_\mathrm{h}(x) &=& \sum_{n=0,1,2,3} c_n \exp( \rho_n x)\\
\rho_n &=& \exp(\mathrm{ i} \theta/4 + \mathrm{i} n \pi/2)
\end{eqnarray} 
The $\rho_n$ have real and imaginary parts of
similar order, so that the solutions decay or increase and have spatial
oscillations on a similar lengthscale, \trp{which is} approximately $\xi$ in
\trp{dimensional} units.   
Due to the asymptotic matching conditions, only
the exponentials that decrease as $x$ increase are allowed,
corresponding to $\rho_0$ and $\rho_3$ if $0 \leq \theta < 2 \pi$.  

\hcf{At the proximal end, due to
the clamped boundary conditions the homogeneous solution must cancel the
inhomogeneous solution at the attachment to the head.  The homogeneous
solution is specified by}
\begin{eqnarray}
c_0 &=& \frac{- \rho_3 h_\mathrm{p}(0) + \mathrm{i} k h_\mathrm{p}(0)}{\rho_3 - \rho_0}\\
c_3 &=& \frac{\rho_0 h_\mathrm{p}(0) - \mathrm{i} k h_\mathrm{p}(0)}{\rho_3 - \rho_0}
\end{eqnarray}
which has an amplitude of order $f_\mathrm{m} (\xi/L)^4 $, the same as the inhomogeneous solution.
\hcf{At the distal end, the largest term in the boundary conditions comes
from balancing the sliding force.  The homogeneous solution is specified by}
\begin{eqnarray}
c_0 &=& \frac{- \rho_3 (\mathrm{i} k)^2 h_\mathrm{p}(1) - [\xi^3 f_\mathrm{m}(1) - (\mathrm{i} k)^3 h_\mathrm{p}(1)]}{\rho_0^2 (\rho_3 - \rho_0)}\\
c_3 &=& \frac{\rho_0 (\mathrm{i} k)^2 h_\mathrm{p}(1) + [\xi^3 f_\mathrm{m}(1) - (\mathrm{i} k)^3 h_\mathrm{p}(1)]}{\rho_3^2 (\rho_3 - \rho_0)}
\end{eqnarray}
which has amplitude of order $f_\mathrm{m} (\xi/L)^3 $.  For high viscosities and low to
moderate Deborah number this \trp{amplitude} can be much larger than the \trp{amplitude of the} proximal and
inhomogeneous solutions.  In the viscous case, with $\mathrm{De} = 0$,
and high viscosity, the ratio between the amplitude of the homogeneous and
inhomogeneous solution in the distal portion of the flagellum is of
order $k/\mathrm{Sp}$. 

\trp{To summarize,} in the middle of
the flagellum the homogeneous solution vanishes and the beating shape
is dominated by the inhomogeneous solution, which has a magnitude
which decreases as viscosity increases, and is quite small for high
viscosity.  At the proximal end of the flagellum, the homogeneous
solution is of the same order as the inhomogeneous solution due to the
clamped boundary conditions.  On the other hand, the motion at the
distal end is dominated by the homogeneous
solution, has larger amplitude than the middle portion, decays
exponentially with a lengthscale $\xi$, and can have oscillations with
wavelengths up to a few times smaller than $\xi$.   
\trp{Thus,} in high viscosity solutions the motion of flagellar beating patterns are
constrained to the distal tip.  Although we assume only a single mode
of the sliding force, the asymptotic analysis is 
general\trp{ly valid}, and should also
apply to more realistic sliding forces.

In Fig.~\ref{ExptShapes}b we plot flagellar beating patterns for a
sliding force of $k=3 \pi/L$.    We constrain the motion to lie
in a two-dimensional plane for simplicity, although the observed
beating patterns are three-dimensional.  These beating patterns are obtained by
solving the linear Eq.~(\ref{NDActive}) \trp{exactly,} 
i.e., not using the asymptotic
analysis.  In this modeling we assume
the pure Hanks' solution and Hanks' solution with polyvinylpyrrolidone
to be Newtonian.   For the Hanks' solution with methylcellulose we
assume a Deborah number of 0.5, corresponding to a beating frequency
of 7\,Hz~\cite{Ishijima1986} and a relaxation time constant of about
$10^{-2}$\,s~\cite{AmariNakamura1973}.  For the
cervical mucus we assume a Deborah number of 50, corresponding to a
beating frequency of 12\,Hz~\cite{Ishijima1986} and a relaxation time constant
of a little less than a second~\cite{TamKatzBerger1980}.  To determine
$\mathrm{De}_2$ we use the ratio $\eta_s /\eta = $1\,cP/4360\,cP.  The
Sperm numbers are determined from $L = 40$\,$\mu$m, $\zeta_\perp = 2
\eta$, $A= 4 \times 10^{-22} \,\mathrm{Nm}^2$~\cite{camalet_et_al1999},
viscosities from Fig.~\ref{ExptShapes}, and frequencies of 12\,Hz for
the 1cP and  35\,cP media, 7\,Hz for the 4000\,cP media, and 12\,Hz for
cervical mucus~\cite{Ishijima1986}. 

In the plots, the amplitude of the beating pattern is proportional to
the driving force $\tilde f_\mathrm{m}$.  The amplitude in the middle sections is
therefore nearly completely suppressed as viscosity increases as
compared to the amplitude in the 1\,cP case.  At the same time, the
motion near the end of the flagellum is relatively unsuppresed.  
In addition the observed beating pattern in cervical mucus has
motion extending farther away from the distal end than the beating
pattern in 4000 cP Hanks' solution, in both the experimental
(Fig.~\ref{ExptShapes}a) and modeled
(Fig.~\ref{ExptShapes}b) beating patterns.  
This agrees with the fact that
the lengthscale $\xi$ decreases as viscoelastic effects become important.

\section{Discussion}

In a viscoelastic medium the forces exerted on a flexible swimmer are
different from those exerted by a Newtonian medium.  
We have
calculated the viscoelastic forces \trp{for a medium with fading memory using resistive force theory.}
 The effects of
hydrodynamic forces on flexible swimmers can be seen in the passive
filament model we have described.  
We have also used a simple model of an active filament to shed some light
on the motion of sperm flagella in different media.  

In agreement with experimental observations, our sliding filament model shows that
the beating is confined to the distal tip for the high viscosity
Hanks' solution.    
\hcf{While the experimental beating patterns are qualitatively explained
by the high-viscosity asymptotic analysis, and the solutions in Fig.~\ref{ExptShapes}b show the same qualitative
trends, our model is not quantitatively accurate.}
For example, in the cervical mucus, our model shows much less
confinement of motion to the distal tip than the observed beating patterns.
In particular, we note that according to our modeling, only the third panel of
Fig.~\ref{ExptShapes}b is in the strongly asymptotic regime, in
contrast to the beating patterns of Ishijima et al, in which the
motion in cervical mucus also seems to be in the asymptotic regime.  
  The discrepancy between our models and the observed
beating patterns may be due to an overestimation of the Deborah number,
our approximation of small amplitudes (the actual beating
patterns show rather large curvatures), or our use of a simple form of the sliding force.   Ishijima et. al. observe varying beating frequencies in different
media\hcf{, indicating} that the sliding force itself is changing in
response to the different drag forces exerted in different media.  
Changes in the active force due to changes in the
media have been modeled by various workers
~\cite{brokaw1971,Lindemann1994, Riedel-Kruse2007}.  We
have specified a fixed force;  however, our analysis might be useful
in extending studies of the sliding mechanism such as in
Ref.~\cite{Riedel-Kruse2007} to high viscosity and viscoelastic situations.
\hcf{Finally, our simple model does not
take into account the possibility of non-Newtonian behavior which
extends beyond viscoelasticity in Hanks' medium with added polymers.}  
We note that in the case of methylcellulose solutions, the swimming of
bacteria with helical flagella has been analyzed by introducing different viscosities for parallel and
perpendicular drag coefficients~\cite{MagariyamaKudo2002}.  This type
of treatment would not
affect our results, since at first order the shape is determined
solely by the perpendicular drag component in our model.  However,
\trp{the work of reference~\cite{MagariyamaKudo2002} points} out that care should be taken to measure appropriate
viscosities for flagellar movements, especially as the viscosity of
methycellulose solutions is known to be shear-rate-dependent.

\trp{Our} work is a first step to understanding how swimming is modified in
a complex medium.  There is scope to expand these studies, probably
numerically, in addressing questions such as the role of end effects
and associated elongational flows, the role of drag from a head
and the role of large amplitude motion in viscolastic fluids.  
In addition, many biological media, including
mucus in the reproductive tract, are in 
gels, which may be
expected to have different effects on the swimming than fluids.

This work was supported in part by National Science Foundation Grants Nos. NIRT-0404031 (TRP) and DMS-0615919 (TRP), and NIH grant No. R01 GM072004 (CWW).
TRP and HCF thank the Aspen Center for Physics, where some of this work was done. 


\end{document}